\documentclass{aa}

\usepackage[varg]{txfonts}
\usepackage{graphicx}
\usepackage{amsmath}

\usepackage[switch,columnwise]{lineno}

\DeclareUnicodeCharacter{2212}{-}
\makeatletter
\renewcommand*\aa@pageof{, page \thepage{} of \pageref*{LastPage}}
\makeatother

\begin{document} 

   \title{Determining the acceleration regions of in situ electrons using remote radio and X-ray observations}
	\titlerunning{Remote and in situ observations of energetic electrons}

   \author{D.~E.~Morosan \inst{1,2}
        \and
        N. Dresing \inst{1}
        \and
        C.~Palmroos \inst{1}
        \and
        J. Gieseler \inst{1} 
        \and
        I.~C.~Jebaraj \inst{1}
        \and
        A. Warmuth \inst{3}
        \and
        A. Fedeli \inst{1}
        \and
        S. Normo \inst{1}
        \and
        J. Pomoell \inst{4}
        \and
        E. K. J. Kilpua \inst{4}
        \and
        P. Zucca \inst{5}
        \and
        B. Dabrowski  \inst{6}
        \and 
        A. Krankowski  \inst{6}
        \and
        G. Mann \inst{3}
        \and
        C. Vocks \inst{3}
        \and
        R. Vainio \inst{1}
        }

   \institute{Department of Physics and Astronomy, University of Turku, 20014, Turku, Finland \\
              \email{diana.morosan@utu.fi}
        \and
             Turku Collegium for Science, Medicine and Technology, University of Turku, 20014, Turku, Finland
        \and
             Leibniz Institute for Astrophysics Potsdam (AIP), An der Sternwarte 16, 14482 Potsdam, Germany
        \and
        Department of Physics, University of Helsinki, P.O. Box 64, FI-00014 Helsinki, Finland
        \and
        ASTRON - the Netherlands Institute for Radio Astronomy, Oude Hoogeveensedijk 4, 7991 PD Dwingeloo, the Netherlands
        \and
        Space Radio-Diagnostics Research Centre, University of Warmia and Mazury, R. Prawochenskiego 9, 10-719 Olsztyn, Poland
        }

   \date{Received ; accepted }

 
  \abstract
    {Solar energetic particles in the heliosphere are produced by flaring processes on the Sun or shocks driven by coronal mass ejections. These particles are regularly detected remotely as electromagnetic radiation (X-rays or radio emission), which they generate through various processes, or in situ by spacecraft monitoring the Sun and the heliosphere. }
    {We aim to combine remote-sensing and in situ observations of energetic electrons to determine the origin and acceleration mechanism of these particles. }
    {Here, we investigate the acceleration location, escape, and propagation directions of electron beams producing radio bursts observed with the Low Frequency Array (LOFAR), hard X-ray (HXR) emission and, in situ electrons observed at Solar Orbiter (SolO) on 3 October 2023. These observations are combined with a three-dimensional (3D) representation of the electron acceleration locations and results from a magneto-hydrodynamic (MHD) model of the solar corona in order to investigate the origin and connectivity of electrons observed remotely at the Sun to in situ electrons. }
    {We observed a type II radio burst with good connectivity to SolO, where a significant electron event was detected. However, type III radio bursts and Hard X-rays were also observed co-temporally with the elctron event but likely connected to SolO by different far-sided field lines. The injection times of the SolO electrons are simultaneous with both the onset of the type II radio burst, the group of type III bursts and the presence of a second HXR peak, however, the most direct connection to SolO is that of the type II burst location. The in situ electron spectra point to shock acceleration of electrons with a short-term connection to the source region.}
    {We propose that there are two contributions to the SolO electron fluxes based on the results and magnetic connectivity determined from remote-sensing data: a smaller flare contribution from the far-side of the Sun and a main shock contribution from the region close to the eastern limb as viewed from Earth. We note that these two electron acceleration regions are distinct and separated by a large distance and connect via two separate field lines to SolO.}


\maketitle


\section{Introduction}

{Solar energetic particles (SEPs) are often observed in the heliosphere with energies ranging from a few keV up to the GeV range. Energetic electrons are of particular interest because they emit electromagnetic radiation, enabling remote observations and allowing us to infer their acceleration mechanisms. This understanding is further enhanced when these electrons are also measured in situ \citep[e.g.,][]{jebaraj2023, morosan2024}, as it provides insights into the transit-time effects influencing their transport from the source to the observer.}

\begin{figure*}[ht]
    \centering
    \includegraphics[width=0.99\linewidth]{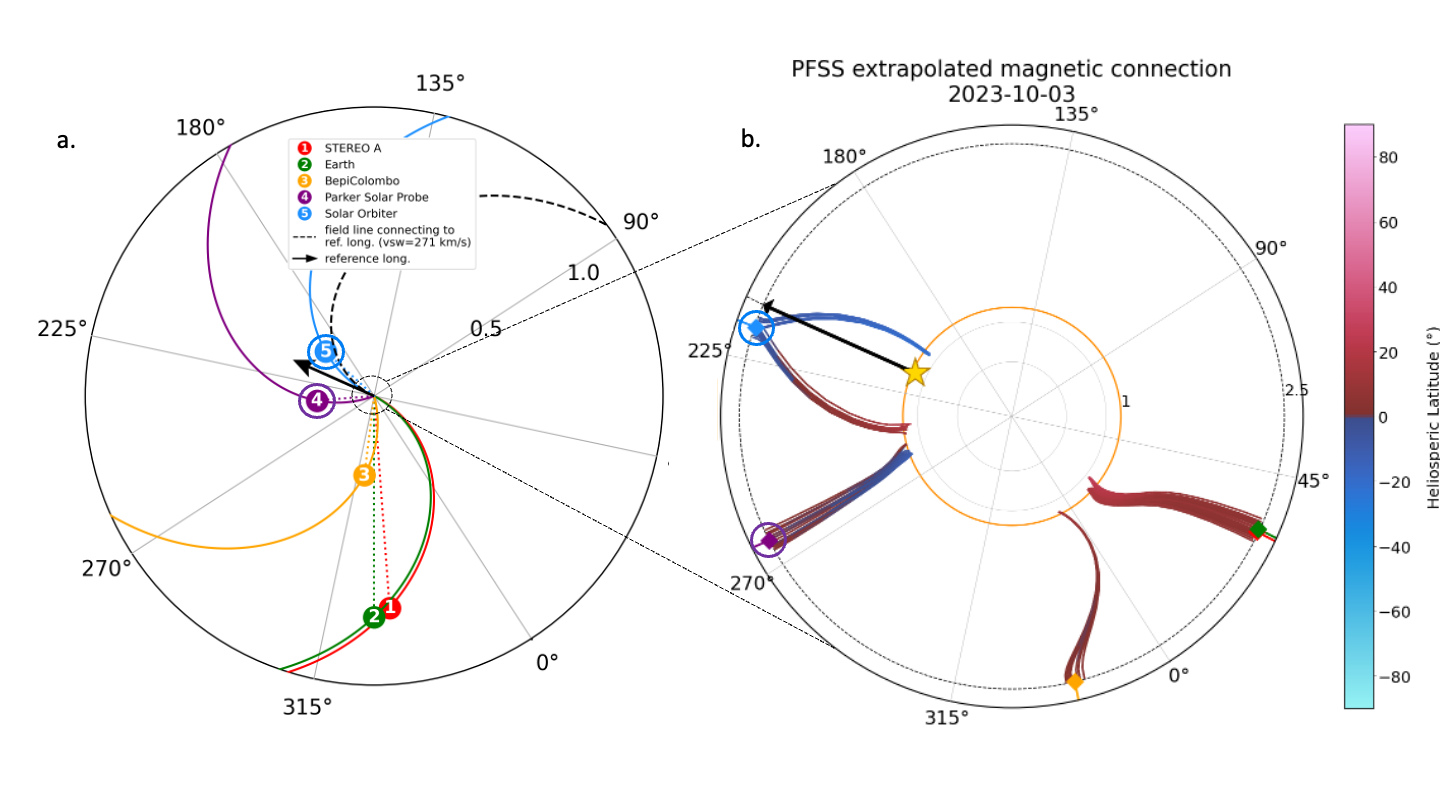}
    \caption{The location of relevant spacecraft monitoring the Sun on 3 October 2023 at 12:00~UT. (a) The Solar-MACH plot \citep[][]{gieseler2023} shows the spacecraft constellation observing the Sun on 3 October 2023. The only spacecraft that observed the in situ particles associated with the far-side CME (SolO and PSP) are differentiated by clear circles. The arrow denotes the direction pointing outwards from the flare site and the dashed spiral line represents the field line connecting to the flare. (b) Zoomed-in view close to the Sun where the coloured solid diamonds mark the magnetic footpoint of each spacecraft on the source surface before connecting to the solar surface via PFSS magnetic field extrapolations. The yellow star denotes the flare location. The colour of the magnetic field lines correspond to heliographic latitude. Two distinct field line bundles are likely to connect to SolO below the source surface.}
    \label{fig:fig1}
\end{figure*}

{Initially, solar flares were considered to be the main mechanism producing solar energetic electrons \citep{lin1982, Reames1999}. Electrons energized in the flaring process travel both downward and upward from the acceleration site along the magnetic field lines. Those electrons that propagate toward the dense chromosphere generate hard X-ray (HXR) emission through the bremsstrahlung mechanism \citep[e.g.,][]{krucker2011}. Those electrons that propagate outward frequently manifest as type III radio bursts generated via plasma instabilities and are co-temporal with the HXR emission, relating both the upward and downward propagating electron populations following flare acceleration \citep[e.g.,][]{vilmer2002, glesener2012, reid2014}. }

{Flares are often associated with coronal mass ejections (CMEs), which can also energize electrons whose signatures are observed concurrently with those related to flare-accelerated electrons \citep[e.g.,][]{jebaraj2023,morosan2024}. These electrons are accelerated by quasi-perpendicular CME-driven shock waves to energies up to a few tens of keV \citep[][]{mann05,mann2022}. They can produce plasma emission \citep[e.g.,][]{wild1950, nelson1985, kl02} that can be detected remotely as type II radio bursts \citep[e.g.,][]{ma96,ne85}. Type II bursts are composed of numerous fine structures, the most well-known being herringbones, which are signatures of individual electron beams escaping the shock \citep[e.g.][]{holman1983,ca87,mann2018, mo19a, magdalenic20}. Type II bursts, generated by CME-driven shock waves, follow the expansion direction of these shocks \citep[e.g.,][]{zimovets2012, zu18, mancuso19, mo19a, Morosan2020a, jebaraj2021}. \citet{morosan2022} showed that herringbone bursts at frequencies above 150~MHz emanate from regions of exclusively closed field lines during the early stages of the CME eruption. However, a more recent study of a different event showed that, depending on the expansion direction of the CME, herringbones can also reach open field lines, further away from the eruption region \citep[][]{morosan2024}. The escaping radio emitting electrons in the study of \citet[][]{morosan2024} intersected magnetic field lines that were well-connected to spacecraft at 1~AU, that also observed a co-temporal electron event. Therefore, radio imaging can effectively demonstrate the role of shock acceleration in the electrons observed in situ by connecting the trajectories of the imaged type II radio sources with interplanetary magnetic field lines. Energetic electrons observed by near-Sun spacecraft have yet to be studied in conjunction with radio imaging of type II bursts. This is largely due to unfavorable spacecraft constellations relative to Earth, where most radio imaging is conducted, as well as the limited availability of radio imaging observations.   }
 
{While radio observations of strong eruptive events on the Sun are common, inferring the relation of in situ particles to the distinct sources of radio emission in dynamic spectra (such as type II and type III bursts) and to HXR emissions has been a long standing issue  \citep[e.g.,][]{gopalswamy2006coronal, Kahler2007, Agueda2009, Kahler2019}. A direct association between in situ electrons, radio bursts and HXRs has not yet been made. Most studies rely on co-temporal occurrences in order to make a link \citep[e.g.,][]{li2021}, while a detailed magnetic connectivity analysis to electron acceleration regions in the low corona has only been attempted recently. For example, \citet{klassen2011} found a series of type III radio bursts to be temporally associated with a series of electron spikes in the low energy range (below 120~keV) detected directly by spacecraft, these two phenomena in turn co-temporal with coronal jets. A recent study by \citet{jebaraj2023} found relativistic electrons in situ that could be related both to a flare and a coronal shock wave, with the flare contribution supported by HXR pulses and the shock contribution by the presence of herringbone bursts. Making the connection between distinct sources of radio emission and their connectivity to electron events at spacecraft has only recently been done by \citet[][]{morosan2024} to determine the possible sources of in situ electrons.}

{In this paper, we present a detailed analysis of an electron event detected at Solar Orbiter \citep[SolO;][]{muller2020} that was associated with an eruption behind the Earth-facing solar limb. We use radio and X-ray observations to distinguish between possible flare and shock contributions to the electron fluxes observed. In Sect.~\ref{sec:analysis}, we give an overview of the observations and data analysis techniques used. In Sect.~\ref{sec:results}, we present the results, which are further discussed in Sect.~\ref{sec:discussion}, where we also present our conclusions.}


\section{Observations and data analysis} \label{sec:analysis}

\begin{figure*}[ht]
    \centering
    \includegraphics[width=0.85\linewidth]{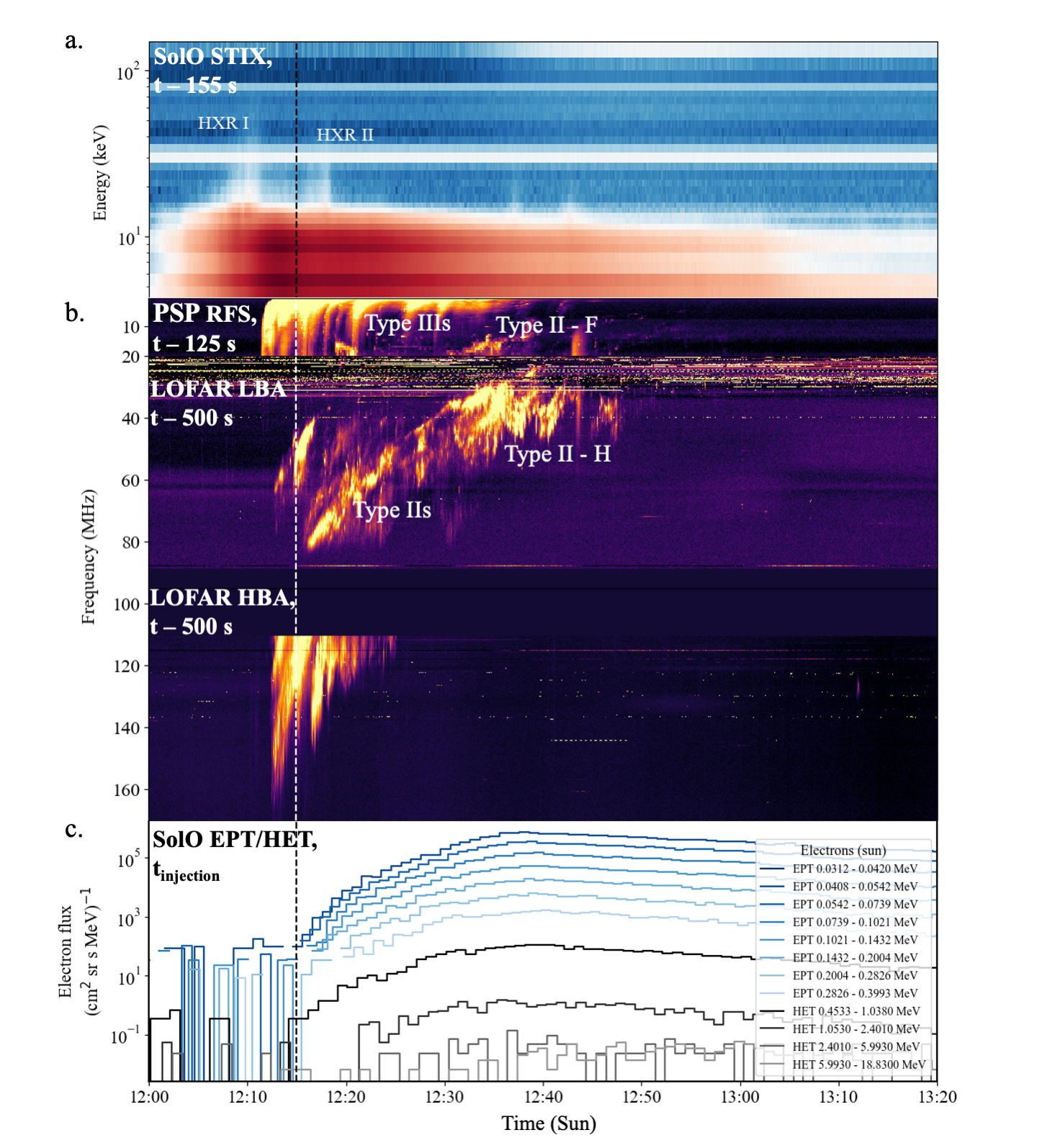}
    \caption{X-ray spectrogram, composite dynamic spectrum of a complex radio event and energetic electron time series observed on 3 October 2023. (a) X-ray spectrogram from SolO/STIX showing the onset of the flare and the presence of two hard X-ray peaks: HXR I and HXR II. (b) The dynamic spectrum consists of spectra from PSP/RFS (1-20 MHz), LOFAR LBA (10--90~MHz) and LOFAR HBA (110--170~MHz). The labels 'F' and 'H' refer to fundamental and harmonic emissions, respectively. (c) Time-series of energetic electrons measured by SolO EPT \& HET from the sunward direction (where the first-arriving electrons are observed) and at multiple energy channels. The time represents the injection time for each channel. The spectra are time-shifted to correspond to time at the Sun so that they can be compared directly to the electron injection times.}
    \label{fig:fig2}
\end{figure*}

\begin{figure*}[ht]
\centering
    \includegraphics[width=0.9\linewidth]{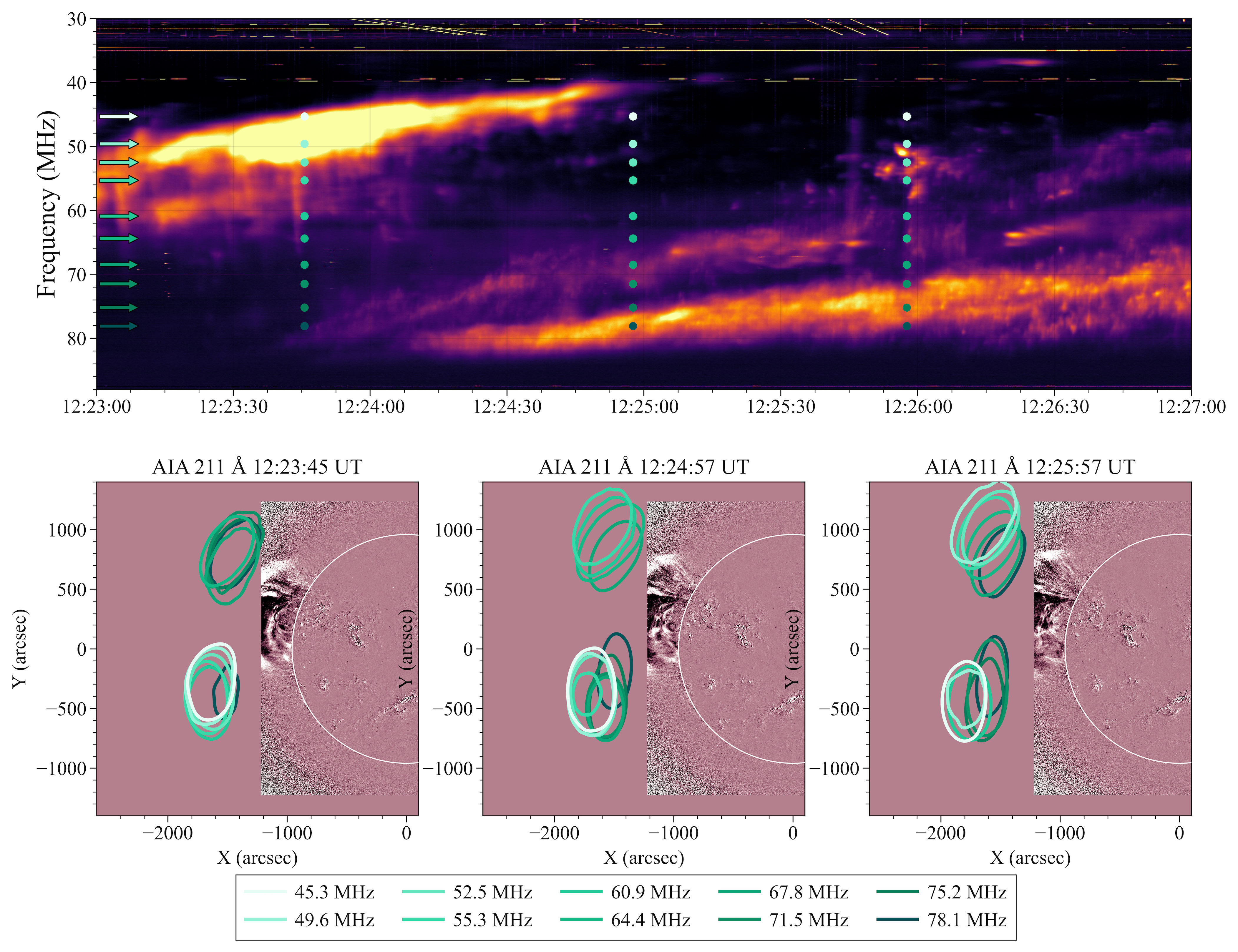}
    \caption{The type II burst in dynamic spectrum and radio images. The top panel shows a zoomed in dynamic spectrum of the onset of the type II harmonic lanes. The arrows located to te left of this panel represent the frequency subbands that were imaged in the bottom panels. The bottom panels show the 70\% radio contours at the time and frequencies denoted in the top panel using the same colouring, overlaid on AIA 211~\AA\ running difference images of the eruption. The radio contours and AIA images have the same time stamps, as shown in the panels. The time labels in this figure are UT observer times.}
    \label{fig:fig3}
\end{figure*}

{On 3 October 2023 at $\sim$12:00~UT solar time, an eruption originating behind the visible solar limb was observed simultaneously with a long-duration ($\sim$40~minutes) type II radio burst recorded with the Low Frequency Array \citep[LOFAR;][]{lofar2013}. This eruption was observed by several space-based observatories at multiple wavelengths. The spacecraft fleet observing the Sun at the time is depicted in the Solar-MACH \citep{gieseler2023} plot in Fig.~\ref{fig:fig1} with each spacecraft or location colour-coded. The only spacecraft that observed the in situ particle event associated with this eruption were: Parker Solar Probe \cite[PSP in purple;][]{psp2016} and SolO (blue).
Fig.~\ref{fig:fig1}a shows the connection between the solar surface to the spacecraft in Carrington coordinates. The black arrow in this figure denotes the flare longitude determined from the mid-point distance between the flare HXR footpoints (see Appendix~\ref{app:a} for more details). Fig.~\ref{fig:fig1}b shows a zoomed-in view close to the Sun where the coloured solid diamonds mark the magnetic footpoint of each spacecraft on the source surface before connecting to the solar surface. The magnetic field lines connecting the spacecraft to the source surface represent the nominal Parker spiral solutions computed considering their heliocentric distances and the observed solar wind speeds. The magnetic connection from the source surface to the solar surface is determined using a Potential Field Source Surface (PFSS) model \citep[][]{Stansby2020} that makes use of a magnetogram from the Global Oscillation Network Group \citep[][]{gong}. This connection is determined tracing a bundle of field lines close to the point on the source surface ballistically connected to the spacecraft. It is important to note that by considering an area around that point rather than the single field line we get an assessment of the uncertainty of the connection. SolO was the only spacecraft with a view of the far side of the Sun while PSP was approximately in quadrature with Earth observing the eastern limb as viewed from Earth (see Fig.~\ref{fig:fig1}). These spacecraft were relatively close to the Sun at distances of 0.31~AU (SolO) and 0.25~AU (PSP). }

\subsection{Remote Observations}

{LOFAR was observing the Sun on 3 October 2023 as part of regular cycle observations (LC20\_001) using two simultaneous modes: interferometric mode \citep[e.g.,][]{mann2018b} and tied-array beams mode \citep[e.g.,][]{morosan2022b}. A long duration type II burst was detected starting at 12:21~UT (Earth time) lasting for $\sim$40~minutes (see Fig.~\ref{fig:fig2}b; however, in this figure all times are shifted to Sun times for comparison with different observatories; the particle times were shifted using the inferred path length estimated in Appendix~\ref{app:c}). The type II was first observed in LOFAR's High Band Antennas (HBAs) starting at a frequency of 170~MHz. The type II continued to drift to low frequencies and was then observed in the Low Band Antennas (LBA). The LBA observations show a complex structure with multiple lanes and fundamental and harmonic emissions (see Fig.~\ref{fig:fig2}b). The majority of the emission in the LBA band is likely harmonic since additional type II lanes were observed by PSP Radio Frequency Spectrometer \citep[RFS;][]{rfs2017} between 10--20~MHz with an approximately 2:1 frequency ratio to the LBA type II lanes (see Fig.~\ref{fig:fig2} where fundamental and harmonic lanes are labelled F and H, respectively). We only used the high-frequency (HFR) spectra from PSP in Fig.~\ref{fig:fig2} since the type II burst does not propagate below $\sim$10~MHz. PSP also observes multiple groups of type III bursts between 12:10--12:30~UT, some likely related to the flare. These type III groups are not observed from the ground by LOFAR. This suggests that the type IIIs are propagating on the far side of the Sun away from the flare site (see Fig.~\ref{fig:fig1} where the nominal magnetic field line connected to the flare is on the far side).  }

\begin{figure*}[ht]
\centering
    \includegraphics[width=\linewidth]{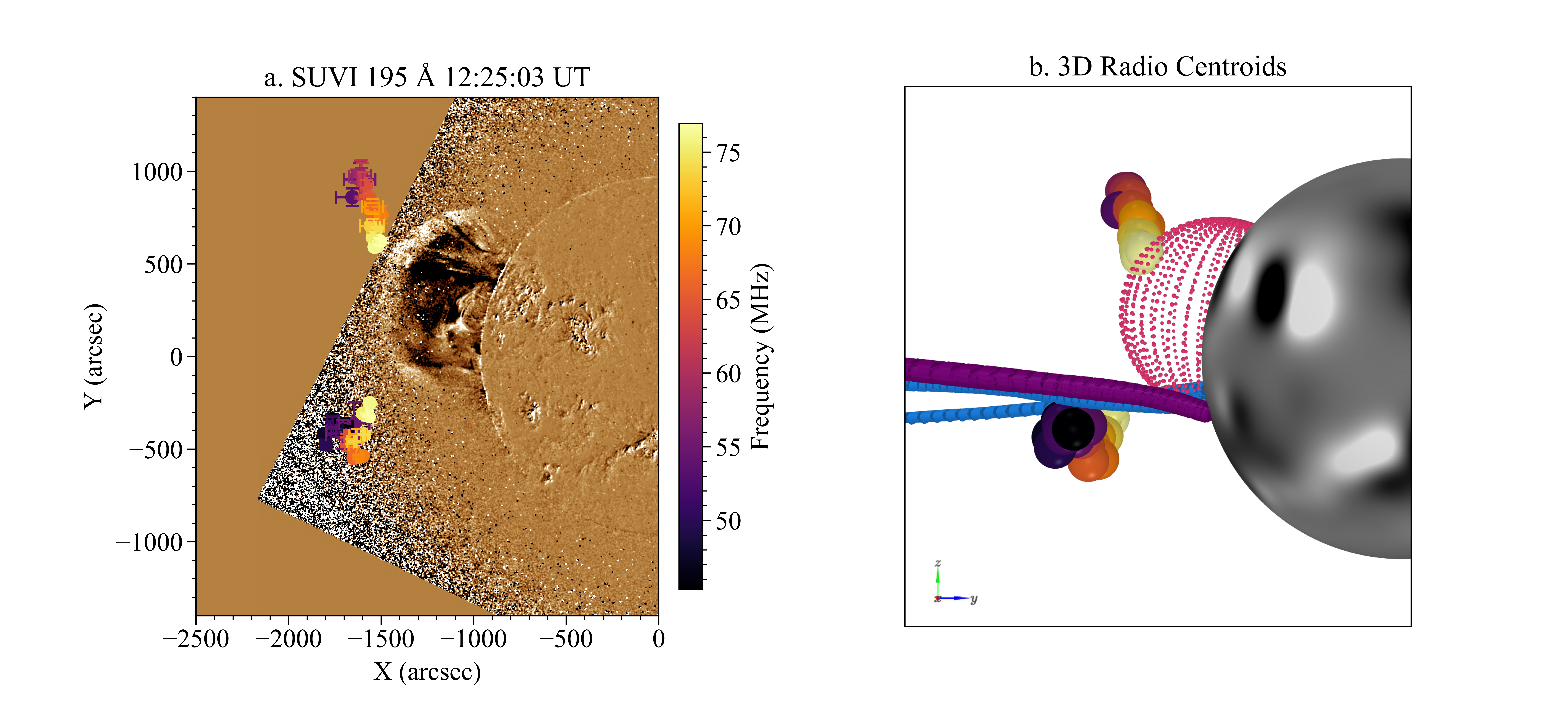}
    \caption{The type II emission centroids and their reconstruction in 3D. (a) Radio source centroids at 12:24:57~UT, where the colour bar denotes the observing frequency in MHz, overlaid on a GOES / SUVI 195~\AA\ image in helioprojective coordinates. (b) Radio source centroids in 3D together with a reconstruction of the CME shock (pink wireframe) and open field lines connecting to SolO (blue) and PSP (purple) in Heliographic Stonyhurst coordinates. }
    \label{fig:fig4}
\end{figure*}

{LOFAR interferometric imaging is available for this observation in the LBA band with 60 frequency sub-bands available for imaging. We used a baseline of up to 15~km in the East-West direction for this study, and the interferometric visibilities were calibrated using the LOFAR Initial Calibration Pipeline \citep[e.g.,][]{linc2019}. The calibrated visibilities were then processed with \texttt{wsclean} \footnote{wsclean \href{https://gitlab.com/aroffringa/wsclean}{https://gitlab.com/aroffringa/wsclean}} \citep{offringaWsclean2014} to produce science-ready images. Due to bad ionospheric conditions, the radio sources and the Quiet Sun were moving in the interferometric images over time periods of minutes. As a result, the images were corrected for ionopsheric refraction by re-centering the Quiet Sun over all the sub-bands for selected time steps when the Quiet Sun was clear in the images. As a result, only the time steps closest to the Quiet Sun images can be corrected, thus, only selected time steps can be chosen to image the type II bursts. The type II lanes are then imaged at several frequencies and their trajectory can be tracked over time. Examples of multi-frequency imaging are shown in Fig.~\ref{fig:fig3} during the onset of the harmonic emissions of the type II lanes (these images include the ionospheric corrections). For the purpose of this study, we only focus on the onset of the type II harmonic emission to compare it to the onset of in situ electrons. A detailed study of the type II burst and its multiple lanes and source locations will be presented in a follow-up paper.  }

{Remote observations also show the onset of a CME on the far-side of the Sun observed by multiple spacecraft such as the Solar and Heliospheric Observatory \citep[SOHO;][]{do95, br95}, Solar Dynamics Observatory \citep[SDO;][]{pe12} and STEREO-A located in orbits near Earth. The first observation of the CME in EUV images was observed in SDO's Atmospheric Imaging Assembly \citep[AIA;][]{le12} images at $\sim$12:23~UT, while in white-light images it was observed at 12:48~UT in the LASCO C2 coronagraph \citep{br95}. SolO was the only spacecraft facing the far side of the Sun, and it observed an eruption originating from a far-side active region with both the EUI \citep[][]{eui2020} and STIX \citep[][]{stix2020} instruments. The flare was accompanied by both hard and soft X-ray emission, and it has a Geostationary Operational Environmental Satellite GOES-flare-class equivalent of C2 (see the STIX spectrogram in Fig.~\ref{fig:fig2}a). The flare in EUV and two accompanying HXR footpoints can be seen in Fig.~\ref{fig:figA1}. No on-disk EUV wave was visible in either EUI or AIA images; however, we note that the cadence of the EUI instrument was low (10 minutes) and the eruption may have been too far behind the limb for a wave to be visible in AIA images. Running difference images of the eruption in the AIA 211-\AA\ channel are shown in Fig.~\ref{fig:fig3} with overlaid multi-frequency contours of the type II burst.}

{The onset and propagation of the eruption as viewed from Earth and the radio bursts associated with it can be seen in running difference images from AIA (Figs.~\ref{fig:fig3} and \ref{fig:fig4}).  }

\begin{figure*}[ht]
\centering
    \includegraphics[width=\linewidth]{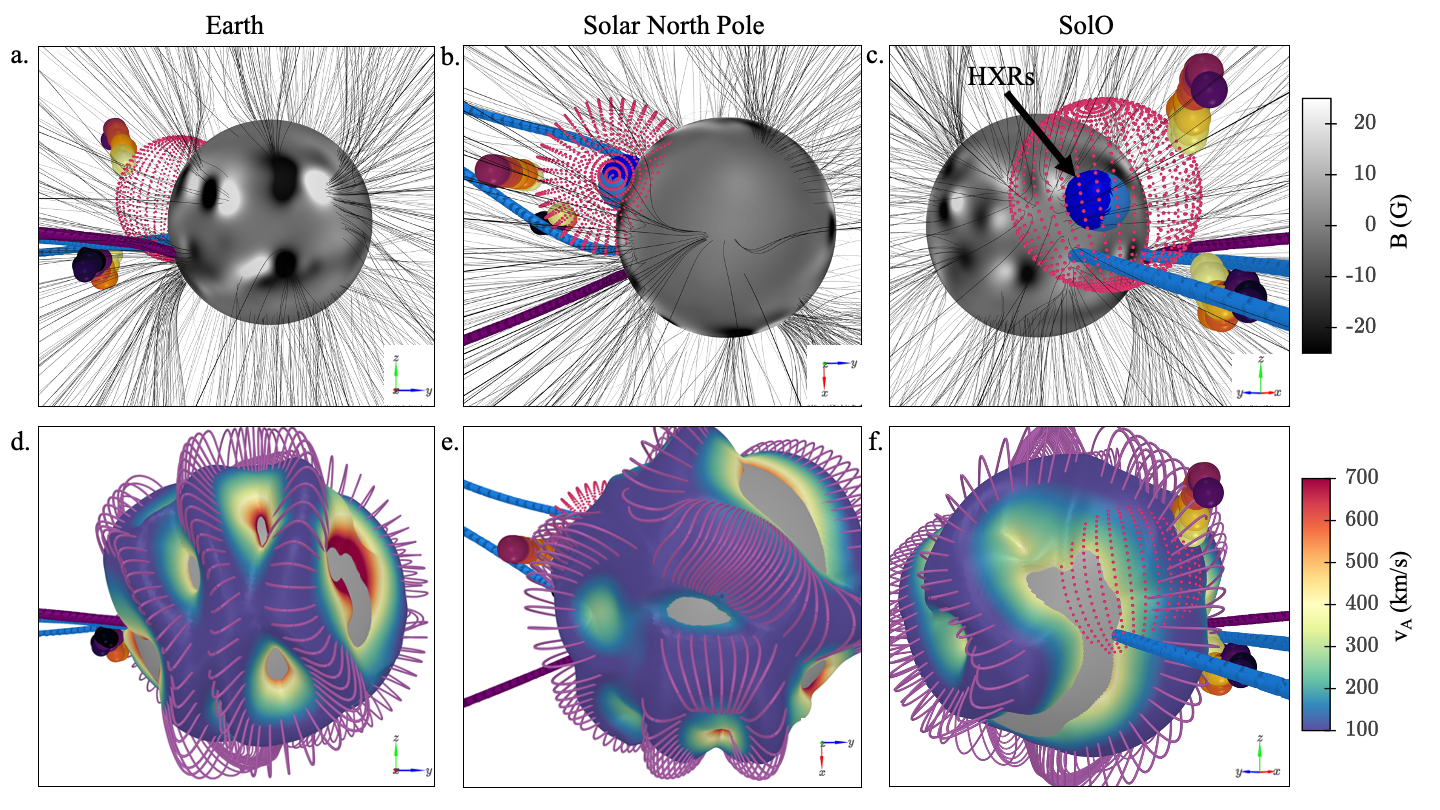}
    \caption{3D locations of the type II radio bursts, HXR footpoints and CME combined with the MHD model results. The locations are shown from three different perspectives: Earth (left), solar North pole (middle), and SolO (right) in Heliographic Stonyhurst coordinates. The axis in each panel shows the orientation of the coordinate system where the x-axis points roughly along the Sun--Earth line. a--c. Radio source centroids in 3D (coloured spheres) together with the HXR footpoints (blue spheres overlaid on the solar surface in panels b and c) and a reconstruction of the CME shock (pink wireframe). These are overlaid on an HMI magnetogram used as input to the MAST model and and open field lines (black) including field lines connecting to SolO (blue) and PSP (purple). d--f. Density iso-surface corresponding to the harmonic plasma frequency level of 76~MHz overlaid with values of Alfv\'en speed at the heights corresponding to the density surface and closed magnetic field lines in purple. The grey sphere underneath the surface denote a radius of 1.4~R$_\odot$. }
    \label{fig:fig5}
\end{figure*}

\subsection{Spacecraft particle observations}

{Energetic electrons and other particles from the Sun were monitored by the spacecraft constellation shown in Fig.~\ref{fig:fig1}. The nominal Parker spiral combined with the PFSS magnetic field from the source surface suggests a magnetic field connectivity which is not distant from the flare site (see Fig.~\ref{fig:fig1}b). SolO observed an intense electron event in both Electron and Proton Telescope \citep[EPT;][]{rodriguez-pacheco2020} and High Energy Telescope \cite[HET;][]{rodriguez-pacheco2020}, shown in Fig.~\ref{fig:fig2}c, and a proton event extending to the highest energy channels of HET, that is, however, not the focus of the present study. The next closest connection to the eruption region is from PSP, which connects close to the eastern limb as seen from Earth. However, PSP only observes a small increase in electrons at energies of 0.4--1~MeV during the type II event, but it does observe a more significant increase in protons compared to electrons (see Appendix~\ref{app:psp}). Other than PSP and SolO, BepiColombo observes both electron and proton events but its magnetic connectivity is on the opposite side of the Sun compared to the eruption site and an analysis of the injection times places these events before the time of the back-sided eruption. Thus, BepiColombo observations are not considered in this study. At STEREO-A and Earth no particle event was observed.   }

\begin{figure*}[ht]
\centering
    \includegraphics[width=0.65\linewidth]{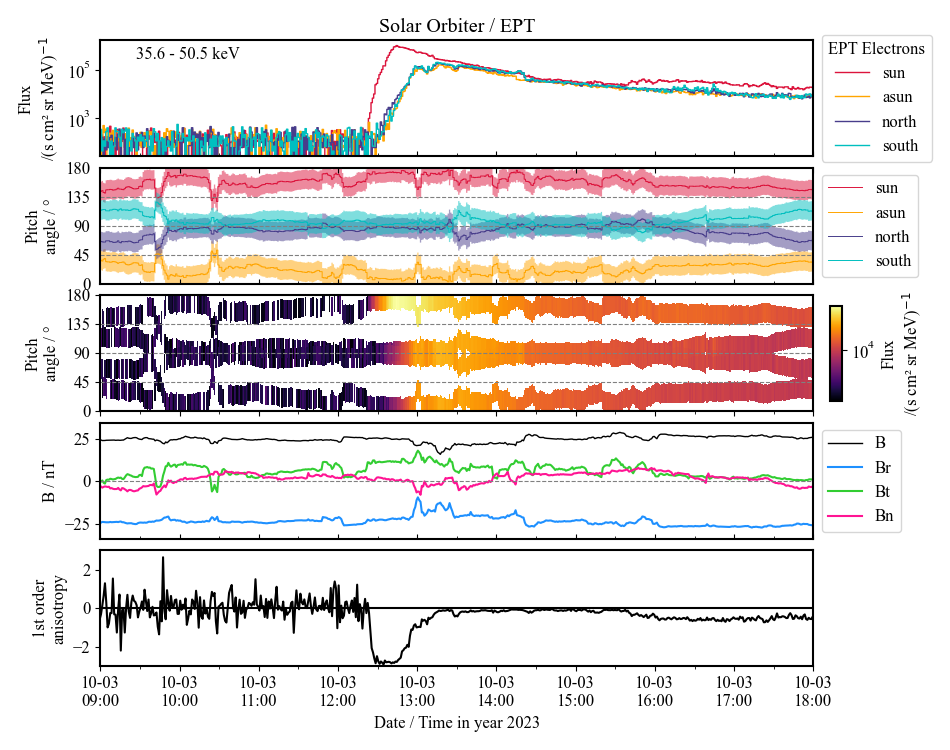}
    \caption{Energetic electrons in the range of 41--54~keV observed in the four viewing directions of SolO/EPT (top panel), and pitch angles covered by the four viewing directions (2nd panel), followed by the colour-coded pitch-angle distribution (PAD), the magnetic field magnitude and RTN-components, and the first-order anisotropy of the electron measurements. We note that the small anisotropy starting around 15:00~UT is not real but due to ion contamination in the SUN sector.}
    \label{fig:fig6}
\end{figure*}


\section{Results} \label{sec:results}

\subsection{Location and characteristics of the radio emission }

{The type II radio burst observed with LOFAR consists of multiple emission lanes. The imaging observations show that there are at least two different locations at the shock front where electrons are accelerated. Fig.~\ref{fig:fig3} shows the position of the type II radio sources at three different times during the onset of the type II lanes and at multiple frequencies. The top panel of Fig.~\ref{fig:fig3} shows a zoomed in dynamic spectrum while the bottom panels show the 70\% radio contours at the time and frequencies denoted in the top panel using the same colouring. These radio contours are overlaid on AIA 211~\AA\ running difference images of the eruption (with a 2~minute running difference period), at the same time as that of the radio contours. The two different type II sources correspond to the expanding northern and southern CME flanks in plane-of-sky images, that are shown in the AIA running difference images in Fig.~\ref{fig:fig3}. The type II sources propagate ahead of the CME in plane-of-sky images. }

{The centroids of the type II sources are extracted by fitting an elliptical Gaussian to each radio source (see for example the methods of \citealt{mo19a}) and all the available frequency subbands in the observation were used. The position of the centroids at 12:24:57~UT are shown in Fig.~\ref{fig:fig4}a, where the colour bar denotes the observing frequency in MHz. In this figure, the centroids are overlaid on a GOES / Solar Ultraviolet Imager \cite[SUVI;][]{suvi2022} 195~\AA\ image, which has a larger field of view than AIA. The entire CME (flux rope and coronal plasma compressed at its leading edge) can be observed in this image before leaving the SUVI field-of-view. In order to de-project the radio source centroids and determine their trajectory in three dimensions (3D), we use the Magnetohydrodynamics Around a Sphere Thermodynamic (MAST) model \citep{lionello2009}, which provides global values of electron density and magnetic field. The MAST electron densities are used to de-project the radio centroids from the plane-of-sky view by comparing the 2D centroid position to the density surface corresponding to the harmonic plasma frequency at each frequency sub-band \citep[see for example the methods of ][]{morosan2024}. The resulting 3D location of the type II bursts is shown in Fig.~\ref{fig:fig4}b from the same view point  as that depicted in Fig.~\ref{fig:fig4}a. The CME shock was reconstructed using the PyThea CME and shock wave analysis tool \citep{Kouloumvakos2022} and is also shown in Fig.~\ref{fig:fig4}b as a pink wireframe mesh. We note that the only available field-of-view for the reconstruction was that from Earth or the nearby STEREO-A spacecraft. Since the CME was behind the limb we made use of the longitude of the eruption estimated from SolO/EUI and STIX images to determine the origin of the shock spheroid. }

{The electron densities combined with the MAST global magnetic field are used to compute the Alfv\'en speed ($v_A$). The results of the MHD model combined with the radio burst locations (Fig.~\ref{fig:fig4}) and position of HXR footpoints (Fig.~\ref{fig:figA1}) are presented in Fig.~\ref{fig:fig5} from three perspectives: Earth (left panels), solar North pole (middle panels) and SolO (right panels), based on the methods of \citet{morosan2022}. The de-projected radio centroids are plotted as coloured spheres for the different type II locations, with the same colouring as in Fig.~\ref{fig:fig4}, which represents frequency. The positions of the HXR footpoints as viewed from SolO are denoted by two overlapping blue spheres in different shades, visible in Figs.~\ref{fig:fig5}b and c. Panels \textit{a--c} of Fig.~\ref{fig:fig5} show the photospheric magnetogram used as input to the MHD model overlaid with open (black) magnetic field lines obtained from the MAST model. Also overlaid in these panels is the reconstructed shock at 12:25~UT shown as a pink wireframe mesh. The HXR footpoint locations (blue spheres) are visible inside this mesh closer to the photosphere. The bottom panels in Fig.~\ref{fig:fig5}, \textit{d--f}, show an electron density iso-surface corresponding to a density layer of $1.8\times10^7$~cm$^{-3}$, which is also the electron density that produces harmonic plasma emission at 76~MHz (the highest frequency of the radio centroids). This density iso-surface at different plasma levels was used to estimate the $z$-component of the plane-of-sky radio centroids so that they can be de-projected from the plane-of-sky view. In panels \textit{d--f}, overlaid on the density iso-surface are the Alfv\'en speed values at the height and location of the density layer of $1.8\times10^7$~cm$^{-3}$ that go from blue (low $v_A\sim$100 km/s) to red (high $v_A\sim$700 km/s). Also overlaid in these panels, are closed magnetic field lines (purple). Both type II sources are located outside but in the vicinity of the reconstructed shock and in regions of enhanced density and low Alfv\'en speed, in agreement with previous studies \citep[e.g,][]{kouloumvakos2021,morosan2024}. The radial plane-of-sky speed of the leading edge of the CME obtained from SUVI images is $\sim$650~km/s indicating that a strong shock was likely to form in the areas where  $v_A\sim$100 km/s. }

\begin{figure*}[ht]
\centering
    \includegraphics[width=0.8\linewidth]{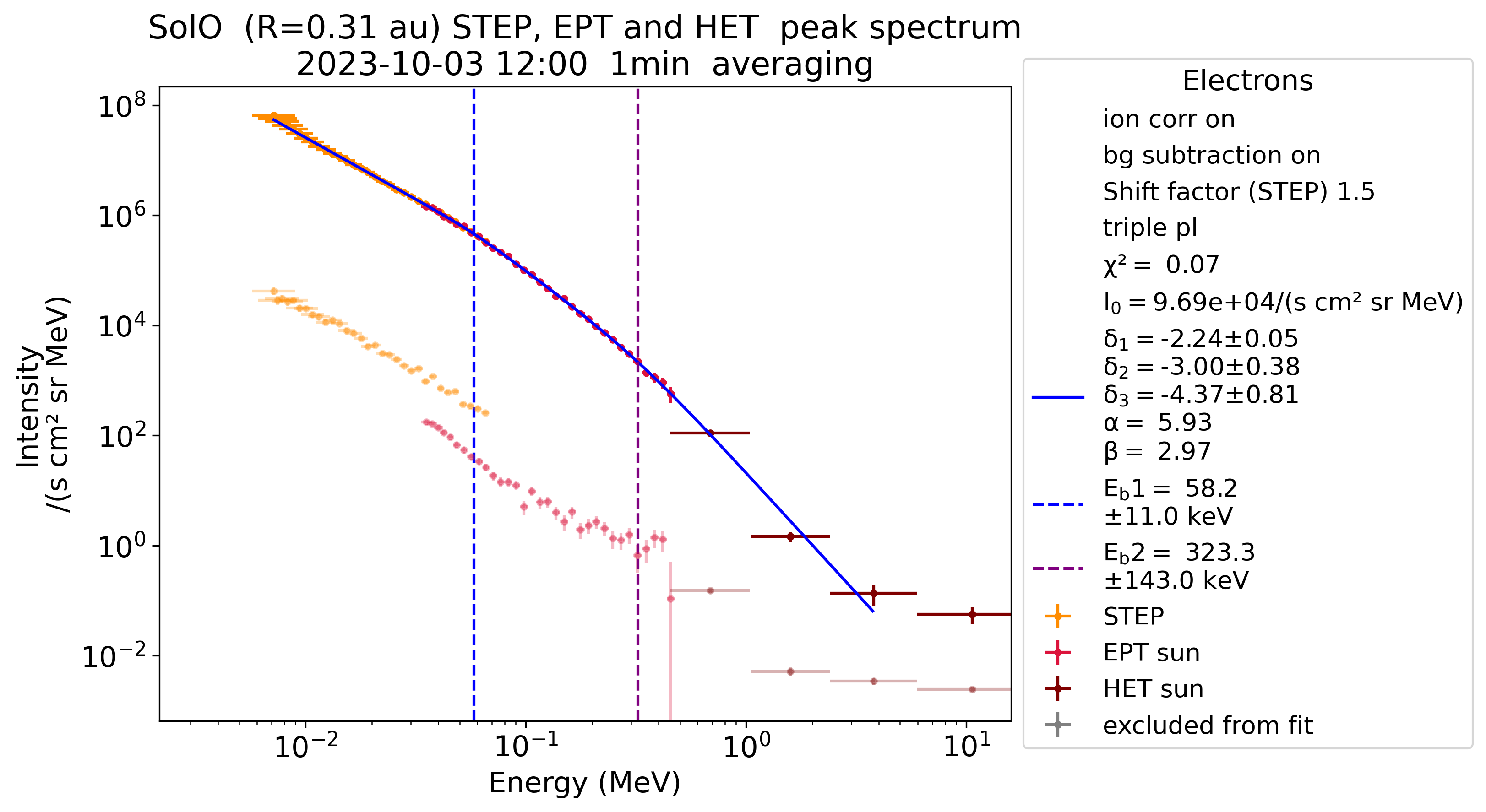}
    \caption{Energetic electron peak intensity spectrum measured by SolO/EPD. The lower and fainter points represent the pre-event background, while the higher points represent the peak intensity measured by STEP, EPT and HET. The spectrum is fit with a triple power-law, excluding the last energy channel of HET.}
    \label{fig:fig7}
\end{figure*}

{In addition, overlaid in the panels of Fig.~\ref{fig:fig5}, are open magnetic field lines that connect to SolO (blue field lines) and PSP (purple field line). These field lines are extracted from the MHD model, and they are in agreement with the connectivity of the PFSS field lines in Fig.~\ref{fig:fig1}. The trajectory of the southern type II source intersects the western extent of the streamer connecting to SolO, that is the field line closer to the visible limb as viewed from Earth. The two blue field lines have opposite polarities, with the field closer to the visible disc being negative (towards the Sun) and the far-side field line being positive. }

\begin{table*}[t]

\caption{Timing analysis between HXRs, radio bursts and in situ electrons.  }
\centering

\label{tab:table1}
\begin{tabular}{|c||c|c|c|c|}
\hline
Event & Onset time  & Emission time  & Energy & Method for     \\
       & at observer (UT) & at the Sun (UT) &   & injection time  \\
\hline
\hline
HXR I       & 12:05             & 12:02              & 15--25~keV              &-   \\
\hline
Type III            & 12:13          & 12:11                 & - & -                 \\
\hline
Type II             & 12:21          & 12:13               & - & -         \\
\hline
HXR II        & 12:17             & 12:14              & 15--25~keV              &-   \\
\hline
Electrons (SolO/EPT+HET)  & 12:22$^*$          & 12:16$\pm 1$ min   & 35-39~keV$^*$ & VDA $L=0.29\pm 0.06$~AU  \\
\hline
\end{tabular}

\tablefoot{The emission time at the Sun for particles represents the injection time estimated using the VDA method, where $L$ is the inferred propagation path length. The emission time at the Sun in the case of the radio bursts and HXRs refers to the time when these are emitted at the Sun taking into account the light-travel time to SolO ($\sim$2~minutes and 5~seconds) and Earth ($\sim$8~minutes and 20~seconds), respectively. The asterisk in the last row indicates the energy ranges at which the SEP onset times have been estimated. The events are ordered chronologically based on their emission time at the Sun.}

\end{table*}

\subsection{Electron onset and injection times}

{The flanks of the CME shock have a direct magnetic connection to SolO based on the reconstruction of the shock and magnetic fields in Fig.~\ref{fig:fig5}. SolO observed an energetic electron event with an onset time of 12:22~UT at energies of 35--39~keV. The electrons were observed up to MeV energies. In order to connect the in situ SEP observations of SolO with the radio and X-ray observations, we infer the electron injection times at the Sun. Electron and proton onset times, inferred injection times, as well as the onset and emission time of the type II bursts and HXR emission, are summarised in Table~\ref{tab:table1}. The emission time in the case of the solar radio bursts represent the times at which the bursts are emitted close to the Sun, taking into account the 1~AU travel time of light. The emission time at the Sun of HXRs takes into account the light-travel time from the Sun to SolO of 125~s. To infer the solar injection times of SEPs, we employed a velocity dispersion analysis \citep[VDA,][]{Lintunen2004}. By assuming a common injection time and propagation path length for all particles, VDA allows for these two parameters to be inferred by using the systematic shift of the arrival times of the SEPs resulting from their different energies. The results of the VDA analysis are shown in Appendix \ref{app:c}.   }

{The closest emission time at the Sun to the injection time of electrons eventually observed by SolO is that of the second HXR peak and the type II radio burst. The emission time of the type II is 12:13~UT while the injection time of electrons is $\sim$12:16~UT. The second HXR peak also occurs at a similar time of 12:14~UT, however, it is less prominent than the first HXR peak. We note that only the southern type II radio burst has a direct connectivity to SolO, while the HXR footpoints are relatively close to the field lines located on the far side that do connect to SolO but do not directly connect to the active region. The type III bursts observed by PSP are not visible in the LOFAR spectra, and therefore no ground-based imaging data is available at higher frequencies. The lack of data does suggests, however, that the type III bursts propagate on the far side of the Sun.}

{Fig.~\ref{fig:fig6} shows directional 41--54~keV electron measurements by SolO/EPT including their pitch-angle distribution (PAD) and the first-order anisotropy (bottom panel), which almost reaches the maximal possible value of $\pm$3. Although the particles are first observed in the sunward facing sector of EPT (SUN), the anisotropy is negative because of the negative polarity of the magnetic field. This magnetic field polarity matches that of the field lines intersecting the type II burst, while the far-side field lines that are closer to the flare have a positive polarity. The very strong anisotropy is consistent with the inferred direct magnetic connection to the electron injection region and suggests that pitch-angle scattering of the electrons during propagation from the Sun to the spacecraft was minor during the early phase of the event. The duration of the anisotropy is rather short, less than an hour, suggesting a short electron injection. However, the time profile of the SUN sector (red trace in the top panel of Fig.~\ref{fig:fig6}) is slightly gradual during the rising phase and less impulsive as would be expected from a very short flare-associated injection. The duration of 15--25~keV X-ray emission (including the HXR peaks) is $\sim$40~minutes (similar to the type II burst duration), however, this emission starts over 10~minutes earlier than the type II burst and the in situ electrons.}

{We also analyzed the electron spectrum observed by SolO as shown in Fig.~\ref{fig:fig7}. The different sensors of SolO/EPD cover a wide electron energy range from tens of keV up to the MeV range. The electron peak spectrum was fitted using the methods described in \citet[][]{Strauss2020} and \citet[][]{Dresing2020}. The fit results in two spectral breaks at $58\pm11$~keV and $323\pm143$~keV with the spectral index evolving from low to high energies of $\delta_1=-2.24\pm0.05$, $\delta_2=-3.0\pm0.38$, and $\delta_3=-4.37\pm0.81$. The reason of spectral breaks in solar energetic electron spectra is still subject to ongoing research. While the source itself could inject a broken power law \citep[e.g., ][]{jebaraj2023b}, the two breaks could also be due to different transport effects in a randomly inhomogeneous plasma \citep[see, e.g.,][]{Kontar2009,Voshchepynets2015, Strauss2020, Dresing2021}.
It has been suggested by \citet[][]{Strauss2020} that the energy range between the two spectral breaks might be the one least changed by transport effects. Comparing the spectral index $\delta_2$ of $-3$ with previous work, we find that it is situated between values expected from flare acceleration, usually resulting in rather soft spectral indices and shocks resulting in harder spectra \citep[e.g.,][]{Oka2018, Dresing2021, Dresing2022}. }

{The electron event appears to be narrow, mainly seen by SolO and only weakly at PSP, which observes the event only in the highest energies. However, the proton event is much broader (seen at both SolO and PSP from low to high energies). This is indicative of a localised source producing energetic electrons such as the flare. However, due to the estimated poor magnetic connection of PSP to the flare site, it is more likely that the electrons observed by PSP were accelerated and injected by the shock. Furthermore, the rather hard spectral index of electrons observed by SolO as well as the slightly gradual rising phase could be indications for a shock contribution to the event seen at SolO. This conclusion is also supported by the presence of a well-connected and long-lasting type II radio burst in the low corona that continues to lower frequencies, indicating continuous shock acceleration.}


\section{Discussion and Conclusion} \label{sec:discussion}

{We presented a complex type II radio burst observed from the ground with LOFAR that occurs simultaneously with a behind-the-limb CME eruption and a far-side strong electron event observed by SolO. The type II radio burst is composed of multiple lanes that are in fact separate bursts coming from  different locations relative to the CME shock. Despite the presence of two distinct type II radio sources at the onset of the electron event, the analysis suggests that only one of them is associated with the source of energetic electrons observed at SolO. Specifically, only the trajectory of the southern type II source intersects one of the field lines connecting to SolO. This type II source is the closest to the ecliptic plane and intersects the field line closer to the visible limb as viewed from Earth. }

{The very strong anisotropy in the SolO electrons proves the very good magnetic connection to the electron injection region. The rather short duration of the anisotropy, less than an hour, does not provide evidence for a sustained electron injection as would be expected from a continuous acceleration by a CME-driven shock. The spectral indices are situated between values expected from flare and shock acceleration. However, it is unlikely that the flare is the main source of energetic electrons observed by SolO due to several arguments: }

{Firstly, the time profile of anti-sunward propagating electrons is slightly gradual during the rising phase and less impulsive compared to that of a typical short flare-associated injection. Secondly, the type II producing region of the shock also has a better connectivity to SolO, compared to the HXR footpoints. The closest emission time at the Sun to the injection time of SolO electrons is indeed that of the type II radio burst, followed by the second X-ray peak (see Table~\ref{tab:table1}). In addition, the in situ magnetic field polarity as SolO is negative, matching the polarity of the field line that intersects the type II burst. The far-side field lines connected to SolO and that are closer to the flare region have instead a positive polarity. Thirdly, the type II duration is $\sim$40~minutes which is close in duration to the rise time of the electron event and the phase of significant anisotropy. And lastly, PSP, which is not at all connected to the flaring region, also observes some electrons at high energies that are most likely due to shock acceleration. The electron and radio observations, therefore, suggest a short (less than an hour) but nevertheless temporally more extended electron injection compared to the duration of the HXR burst or a delta function injection expected for pure flare acceleration. This is indicative of ongoing acceleration by a shock, however, with a temporally limited connection to the acceleration region. }

{We note that the HXR footpoints are relatively close to the field line located on the far side of the Sun that connects to SolO but there is no direct connection to the active region. Significant type III activity is visible only at PSP and SolO, while, in observations at Earth from the Wind spacecraft \citep[][]{bougeret1995}, the type IIIs start at much lower frequencies suggesting that they are partly occulted and originate on the far side of the Sun\footnote{See, for example: \url{https://parker.gsfc.nasa.gov/static_data/crocs/crocs_flux/2023/crocs_flux_20231003_12.png}}. As a result, there is no imaging available of the type III radio bursts at higher frequencies, since LOFAR does not see the occulted higher frequency part of the type III bursts. At lower frequencies, direction finding techniques can be used to determine the propagation directions of the type III bursts, however, at much larger heights than those of the type II in the present study. The most intense type III emission is observed at SolO, indicating that the type IIIs are likely to propagate more towards the SolO direction. It is worth investigating further multi-spacecraft events related to radio bursts, where imaging is possible for both flare-related type IIIs and shock accelerated type II bursts. Using an in-depth magnetic connectivity analysis it may be in the future possible to determine which spacecraft are connected to the type III and type II locations and if these locations are indeed separated. Additionally, acceleration occurs at multiple locations (for example the northern and southern type II bursts) and the escape of energetic particles can be further investigated once the orbit of SolO reaches outside of the ecliptic in the future. The purpose of such observations will be to better distinguish between flare and shock contributions to observed in situ electrons and their properties. }




\begin{acknowledgements}{This study has received funding from the European Union’s (EU's) Horizon 2020 research and innovation programme under grant agreement No.\ 101004159 (SERPENTINE) and EU's Horizon Europe research and innovation programme under grant agreement No.\ 101134999 (SOLER). This research reflects only the authors' view and the European Commission is not responsible for any use that may be made of the information it contains. D.E.M acknowledges the Research Council of Finland project 'SolShocks' (grant number 354409). N.D.\ and I.C.J are grateful for support from the Research Council of Finland (SHOCKSEE, grant No.\ 346902). J.P. acknowledges the Research Council of Finland Project 343581 (SWATCH). The research is performed under the umbrella of the Finnish Centre of Excellence in Research of Sustainable Space (FORESAIL) funded by the Research Council of Finland (grant no. 352847 and 352850). The authors wish to acknowledge CSC -- IT Center for Science, Finland, for computational resources. This paper is based (in part) on data obtained with the LOFAR telescope (LOFAR-ERIC) under project code LC20\_001. LOFAR (van Haarlem et al. 2013) is the Low Frequency Array designed and constructed by ASTRON. It has observing, data processing, and data storage facilities in several countries, that are owned by various parties (each with their own funding sources), and that are collectively operated by the LOFAR European Research Infrastructure Consortium (LOFAR-ERIC) under a joint scientific policy. The LOFAR-ERIC resources have benefited from the following recent major funding sources: CNRS-INSU, Observatoire de Paris and Université d'Orléans, France; BMBF, MIWF-NRW, MPG, Germany; Science Foundation Ireland (SFI), Department of Business, Enterprise and Innovation (DBEI), Ireland; NWO, The Netherlands; The Science and Technology Facilities Council, UK; Ministry of Science and Higher Education, Poland.}\end{acknowledgements}

\bibliographystyle{aa} 
\bibliography{aanda.bib} 

\newpage

\begin{appendix} 

\section{STIX X-ray Imaging} \label{app:a}

\begin{figure}[h]
\centering
    \includegraphics[width=\linewidth]{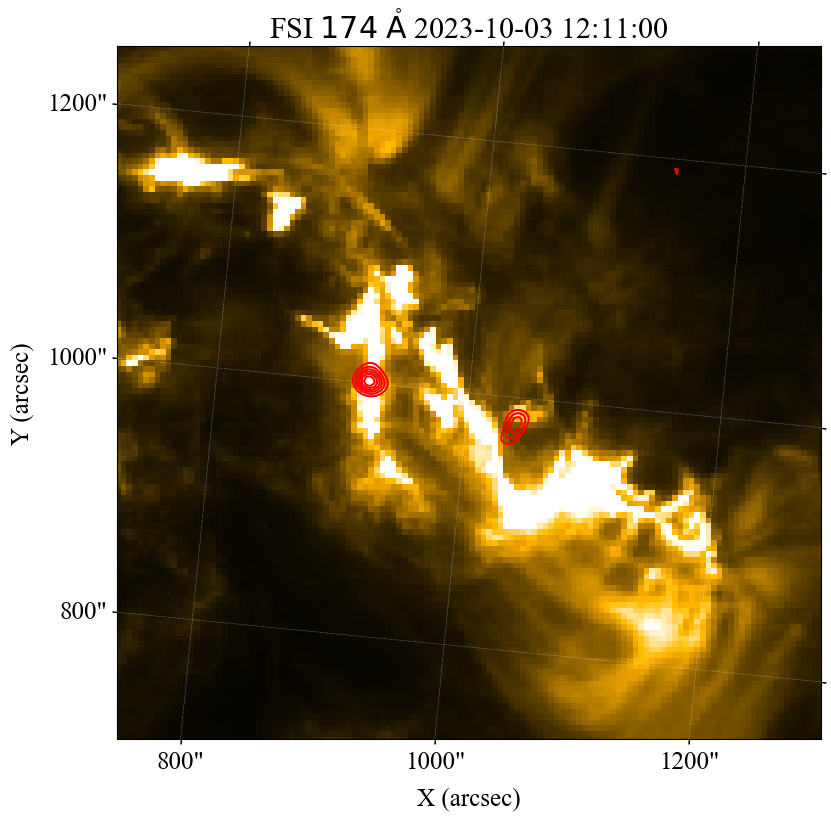}
    \caption{EUI image of the Sun as viewed from SolO. Overlaid on the EUI image are the STIX HXR contours in red from 40-90\% of the maximum intensity level at energies of 20--40~keV and integrated time period of 12:17:30-12:20:30 UT.  }
    \label{fig:figA1}
\end{figure}

{In order to determine the flare longitude and position of the HXR peaks, we produced STIX X-ray images processed with the MEM\_GE algorithm. The STIX X-ray contours at 20--40~keV are presented in Fig.~\ref{fig:figA1} overlaid on a SolO/EUI 174~\AA\ image of the flaring active region. The energy range corresponds to non-thermal emission that represents the two footpoints of the solar flare at 12:12~UT (observer time), before the onset of the second HXR peak. During the second HXR peak, there were not enough counts to reconstruct an image. The flare position in Fig.~\ref{fig:fig1} was estimated by calculating the midpoint between the two HXR footpoints.}

\section{Parker Solar Probe particle observations} \label{app:psp}

\begin{figure*}[h]
\centering
    \includegraphics[width=0.9\linewidth]{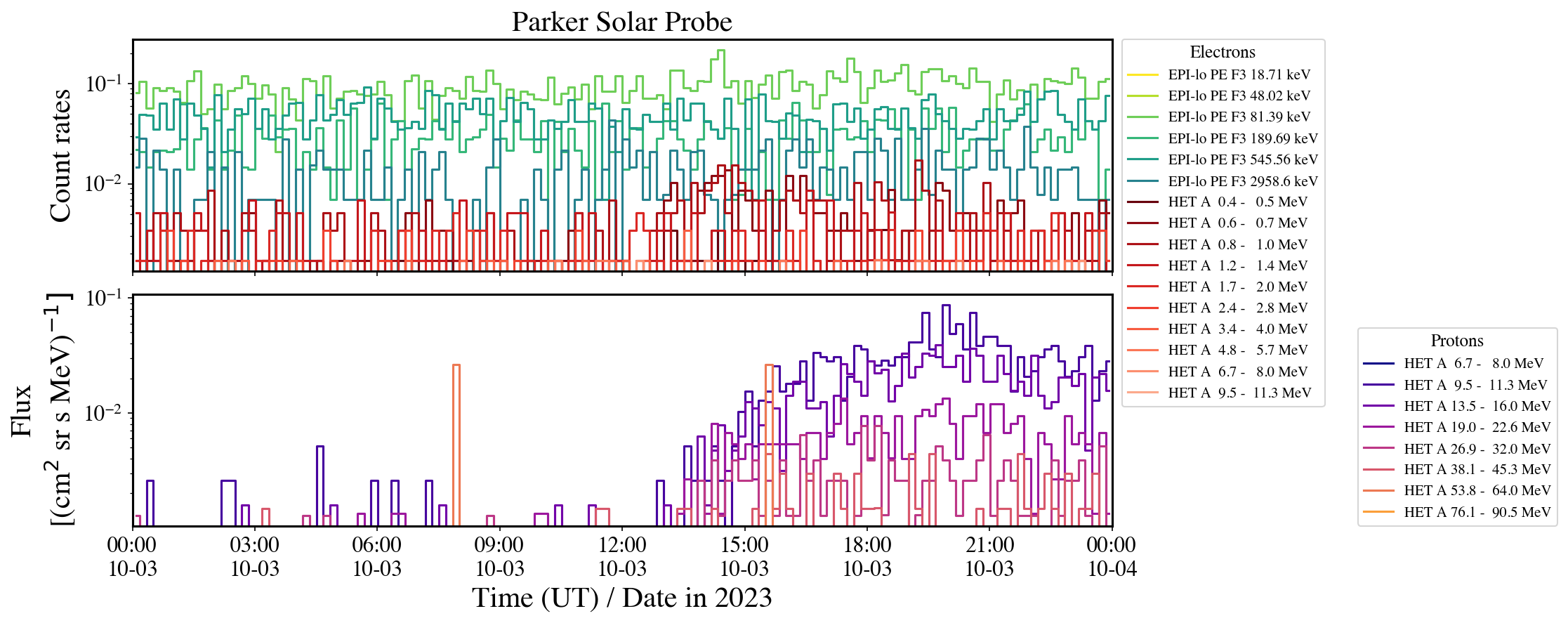}
    \caption{Time-series of energetic electrons (top) and protons (bottom) in different energy channels measured by PSP/IS$\odot$IS. }
    \label{fig:figB1}
\end{figure*}

{Energetic electron and proton observations at PSP in Fig.~\ref{fig:figB1} were obtained by the Integrated Science Investigation of the Sun \citep[IS$\odot$IS;][]{McComas2016} suite. Low-energy electrons were detected in the sunward wedge 3 of channel F of the particle energy (PE) mode of the Energetic Particle Instrument-Low (EPI-Lo), whereas high-energy electrons and protons were provided by the sunward-looking side A of the High Energy Telescope (HET) of the Energetic Particle Instrument-High (EPI-Hi). 
}

\section{Velocity dispersion analysis} \label{app:c}

\begin{figure}[h]
\centering
    \includegraphics[width=\linewidth]{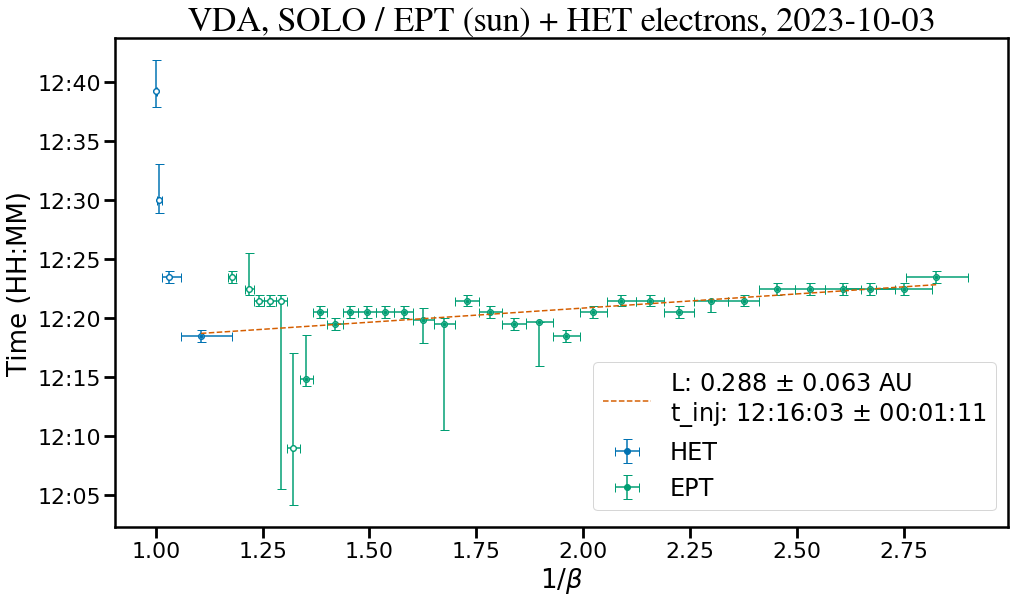}
    \caption{Velocity dispersion analysis plot on SolO EPT+HET electron onset times. Horizontal axis corresponds to the inverse unitless velocity of the electrons ($1/\beta = c/v$) and the vertical axis shows time. Data points mark the onset times, with horizontal error bars displaying the width of each energy channel and vertical error displaying the uncertainty related to the onset times. The dashed orange line shows a linear fit applied on the selected (solid circles) onset times using orthogonal distance regression (ODR) algorithm. Onset times marked with hollow circles are not considered to the fit for their deviance from the general velocity dispersion trend.}
    \label{fig:figC1}
\end{figure}

\begin{figure}[h]
\centering
    \includegraphics[width=\linewidth]{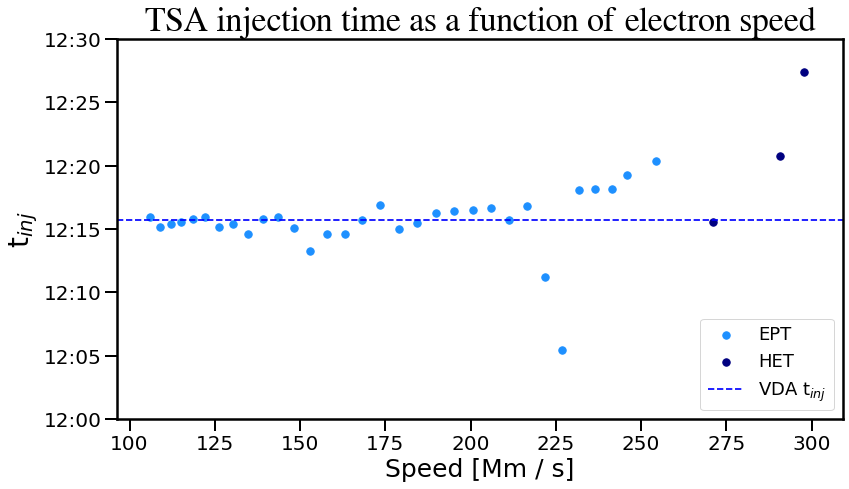}
    \caption{A scatter plot showing the injection times acquired by a simple time-shift ($t_{inj} = t_{onset } - L/v$) for EPT and HET onset times. The horizontal dashed blue line represents the injection time predicted by VDA (Fig.~\ref{fig:figC1}). Largely static injection time as a function of energy, assuming the electrons travelled the same path, suggests a common time of injection.}
    \label{fig:figC2}
\end{figure}

{
To infer the solar injection times of SEPs, we employed velocity dispersion analysis (VDA), that assumes a common injection time and propagation path length of all particles. We performed a VDA for electrons from EPT and HET observed by SolO, leaving out the highest energy channels of both EPT and HET, that were clearly out of the general velocity dispersion trend. The selection criterion was chosen by eye. The results of the electron VDA analysis are shown in Fig.~\ref{fig:figC1}, and they point to an injection time of 12:16 $\pm$ 00:01 and a path length of $L = 0.29 \pm 0.06$ AU. 
Applying time-shift analysis (TSA) on all the individual electron energies, which means simply backtracking to a time of injection from the onset time assuming a path length corresponding to the nominal Parker spiral, we found that the injection times are for the most part constant across energies, and importantly, consistent with the injection time resulting from VDA (Fig.~\ref{fig:figC2}).
}

\end{appendix}

\end{document}